%% this is a Plain TeX file; 5 figures available upon request.  
%% Bruno Nachtergaele, bxn@math.princeton.edu
%
%WORK style file for Proceedings of the Sixth Vilnius Conference
%TEV Publishing Publishing Services Group, Vilnius, 1993
\hsize=15cm
\vsize=25cm
\parindent=0pt

\gdef\thecatchlines{%
\vbox{\noindent\sevenrm Prob. Theory and Math. Stat., Vol. 1, pp.
000 --000\\ B. Grigelionis et al.\ (Eds.)\\
1994 VSP/TEV}}%
\def\topmatter{\tenrm}
\def\endtopmatter{}
\def\\{\hfill\break}
\def\nohyphen{\pretolerance=10000 \tolerance=10000
\hyphenpenalty=10000 \exhyphenpenalty=10000}
\def\title#1\endtitle{\vskip1cm\vbox{\nohyphen\raggedright
\bf\uppercase{#1}}}
\def\shorttitle#1{%\xdef\
}
\long\def\author#1\endauthor{\vskip.5cm#1}
\long\def\address#1\endaddress{\vskip.5cm\vbox{\parskip=6pt#1}}
\long\def\abstract#1{\vskip.5cm\leftline{\bf ABSTRACT}\vskip.25cm#1}

\def\heada#1#2{\vskip.5cm\vbox{\raggedright
\nohyphen\bf\if?#1?\else#1.\ \fi 
\uppercase{#2}}
\vskip.25cm}

\def\headb#1#2{\vskip.5cm\vbox{\raggedright
\nohyphen\if?#1?\else{#1}.\ \fi \sl#2}
\vskip.25cm}

\def\headc#1{
    \smallskip\hskip0.5truecm
    {\sl #1\hskip1truept}}

\def\procl#1#2{\medbreak
\noindent\hskip.5truecm{\sl #1\/}\if?#2?\else\ #2\fi.}

\def\proof#1{\procl{Proof\/}{#1}}
\def\endproof{}

\font\sc=cmcsc10
\long\def\thmo#1#2#3{\medskip\medbreak
    {\hskip0.5truecm\sc #1\if?#2?\else {\ \rm #2}\fi.\enspace}%
    \begingroup\sl #3}
\def\endthmo{\endgroup\par
    \ifdim\lastskip<\medskipamount 
\removelastskip\penalty55\medskip\fi}

\long\def\thm#1 #2\endthm{\thmo{Theorem}{#1}{#2}\endthmo}

\long\def\lem#1 #2\endlem{\thmo{Lemma}{#1}{#2}\endthmo}

\long\def\cor#1 #2\endcor{\thmo{Corollary}{#1}{#2}\endthmo}

\long\def\prop#1 #2\endprop{\thmo{Proposition}{#1}{#2}\endthmo}

\long\def\thmm#1#2 #3\endthmm{\thmo{#1}{#2}{#3}\endthmo}%

%%% REFERENCES 
 
\newdimen\refhangindent 
\refhangindent=.5cm

\def\refs{% 
\vskip.8truecm 
\leftline{\bf REFERENCES}
        \vskip3pt 
        \global\everypar{\hangindent\refhangindent}% 
        \parindent=0pt\rm
\frenchspacing}
\def\book{\expandafter\csname bookr\endcsname}
\def\inbook{\expandafter\csname inbookr\endcsname}
\def\journal{\expandafter\csname journalr\endcsname}
\def\rauthor#1{\rm #1}
\def\rtitle#1{\rm \ #1}
\def\ryear#1{\rm \ (#1).}
\def\rjournal#1{\sl \ #1}

\def\rbook#1{\ \sl#1.}

\def\rpublisher#1{\rm \ #1.}
\def\rvolume#1{\bf\ #1\rm ,}
\def\rjpp#1{\rm \ #1.}

\def\supp#1{\rm \ #1}

\def\ref#1\endref#2{\vskip6pt\def\theref{#1}
\hang\theref\par}

\voffset = -.5in

\input amssym.def % AMS fonts
\input amssym.tex % AMS fonts
% Author's definitions

%for draft only:
%\vsize 23truecm
%\headline={\eightrm Proceedings Vilnius,\  versionof 
%\today \hfill\number\pageno}
%\footline{ }

%general macros
\def\tom{{\tilde{\omega}}}
\def\Sh{{\hat S}}

\def\bond{\vbox to 3pt{\hrule width 1.0truecm height 1pt}}
\def\bondspace{\hskip 1.0truecm}
\def\site{$\hskip 1pt\bullet\hskip 1pt$}
\def\up{\; \uparrow}
\def\down{\; \downarrow}
\def\bg{{\bf g}}

\def\today{\ifcase\month\or
 January\or February\or March\or April\or May\or June\or
 July\or August\or September\or October\or November\or December\fi
 \space\number\day, \number\year}

\font\eightrm=cmr8

\def\caption#1{\centerline{\vtop spread 14pt {
\hsize=13truecm\eightrm #1\hfill}}}

% BLACKBOARD BOLD
\def\idty{{\leavevmode{\rm 1\ifmmode\mkern -5.4mu\else
                                            \kern -.3em\fi I}}}
\def\Ibb #1{ {\rm I\ifmmode\mkern -3.6mu\else\kern -.2em\fi#1}}
\def\Ird{{\hbox{\kern2pt\vbox{\hrule height0pt depth.4pt width5.7pt
    \hbox{\kern-1pt\sevensy\char"36\kern2pt\char"36} \vskip-.2pt
    \hrule height.4pt depth0pt width6pt}}}}
\def\Irs{{\hbox{\kern2pt\vbox{\hrule height0pt depth.34pt width5pt
       \hbox{\kern-1pt\fivesy\char"36\kern1.6pt\char"36} \vskip -.1pt
       \hrule height .34 pt depth 0pt width 5.1 pt}}}}
\def\Ir{{\mathchoice{\Ird}{\Ird}{\Irs}{\Irs} }}
\def\ibb #1{\leavevmode\hbox{\kern.3em\vrule
     height 1.5ex depth -.1ex width .2pt\kern-.3em\rm#1}}
 \def\Cx {{\ibb C}} \def\Rl {{\Ibb R}}

\def\QEDD {\hfill\break
     \line{\hfill{\vrule height 1.8ex width 1.8ex }\quad}
      \vskip 0pt plus100pt}

\def\nl{\hfill\break}

% OPERATORS

\def\A1n{A_1\otimes\cdots\otimes A_n}

  % redefinition!

  % redefinition!
  % \Set\Big#1 to force size of \set

  % In displayed formulas

 % projection onto component

         % used for spectral radius

\def\supp{{\rm supp }}
  % also in text
\def\tover#1#2{{\textstyle{#1\over #2}}}

\def\Tr{\mathop{\rm Tr}\nolimits}

% LETTERS
\def\phi{\varphi}            % redefinition!
\def\epsilon{\varepsilon}    % redefinition!

\def\A{{\cal A}} \def\B{{\cal B}}  
 \def\H{{\cal H}}  
   \def\Z{{\cal Z}}

\def\E{{\Ibb E}}   

\def\d{{\rm d}}
\def\tprod{{\prod}^*}         %time-ordered product
\def\bsig{\sigma\!\!\!\!\sigma}     %boldface sigma
\def\pauli{\bsig} %vector of Pauli matrices

\def\Prob#1{\hbox{\rm Prob}(\hbox{\eightrm #1})}  
%Prob(description of event)
\def\Probin#1#2{\hbox{\rm Prob}_{#1}(\hbox{\eightrm #2})}
%Prob_measure(...)
\def\Ind#1{\hbox{\rm \bf I}[\hbox{\eightrm #1}]}    
%I[description of event]

\def\tom{\tilde{\omega}}
\def\Sh{{\hat S}}
\def\sign{\hbox{sign}}

               % "." is not end of sentence
               % "." is not end of sentence
             % = old  "E hat"

\def\om{\omega}

   % mean entropy

% SU2-related:

\def\es{{\bf S}}

\def\bI{{\rm\bf I}}

\topmatter%-------------------------------------------------  
\thecatchlines \footnote{}{Copyright \copyright\ 1993 by the author. 
Faithful reproduction of this article by any means is permitted for 
non-commercial purposes.}
\title
Quasi-state decompositions for quantum spin sytems
\endtitle
\shorttitle{Quasi-state decompositions for quantum spin sytems}
\author
Bruno~Nachtergaele\footnote{*}{Partially supported by NSF Grants No.
PHY92--14654 and PHY90--19433 A02.}
\endauthor
\address
Department of Physics, Princeton University,\\
P.O. Box 708, Princeton, NJ 08544-0708, USA\\
E-mail: bxn@math.princeton.edu
\endaddress

\endtopmatter%------------- T E X T   B E G I N S --------------- 
\abstract{
I discuss the concept of quasi-state decompositions for ground
states and equilibrium states of quantum spin systems.  Some recent
results on the ground states of a class of one-dimensional quantum
spin models are  summarized and new work in progress is presented. I
also outline some challenging open problems and conjectures.
}

\heada{}{Introduction}

The aim of this contribution is threefold. First I would like to
review the notion of quasi-state decomposition as it was developed
and used in (Aizenman and Nachtergaele, 1993a). I then show how
interesting quasi-state decompositions  for the ground states of a
class quantum spin models can be obtained using a Poisson integral
representation of the Gibbs kernel $e^{-\beta H}$. Secondly I will
try to give a short review of the results we have recently obtained
using quasi-state decompostions. These results mainly concern  a
class of one-dimensional quantum spin Hamiltonians introduced by
Affleck (Affleck, 1985). The third and last part is devoted to a
discussion of work in progress and some open problems in the subject
as well as some stimulating conjectures.

\heada{2}{Quasi-state decompositions and Poisson  integral
representations}

A quantum system is determined by a C*-algebra of observables $\A$
and a one-parameter group of automorphisms of $\A$ representing the
dynamics. Here we are mainly interested in quantum spin systems
defined on a finite lattice $\Gamma$ for which
$\A\equiv\A_\Gamma=\bigotimes_{x\in\Gamma} \A_{\{x\}}$ with
$\A_{\{x\}}\cong M(d,\Cx)$ (the complex $d\times d$ matrices), for
all $x\in\Gamma$. Often $\Gamma$ is a subset of an infinite lattice
(e.g. $\Ir^\nu$) and  one is then interested in properties of the
system as $\Gamma$ tends to the full lattice (the thermodynamic
limit). The dynamics is usually defined in terms of a Hamiltionian
$H=H^*\in\A$ and we will give several examples of interesting
Hamiltonians below. A state of the system is a normalized positive
linear functional of $\A$. A class of physically  interesting states
are the equilibrium states: for any  $\beta\geq 0$ one defines a
state $\omega_\beta$ by $$ \omega_\beta(A)={\Tr A e^{-\beta
H}\over\Tr e^{-\beta H}} \eqno(2.1)$$  for all $A\in\A$. In this
paper by ground state we always mean a state which is a limiting
point of (2.1) as $\beta\to\infty$, possibly under specified
boundary conditions. In most examples discussed here (2.1) will
converge to a unique limit for finite systems and the thermodynamic
limit will be taken afterwards.

It has been remarked many times that analyzing the structure of 
equilibrium states and ground states of a quantum system poses extra
difficulties as compared to the situation in classical equilibrium
statistical mechanics where the states are probability measures.
Sure enough these probability measures can be highly non-trivial as
well, but the fact that there is an underlying configuration space
which can be visualized in a concrete way helps a lot in
understanding their behaviour.

The aim of this section is to explain how at least for some quantum
spin models one can obtain a pictorial representation of the ground
states and equilibrium states which resembles somewhat the situation
for classical systems. More precisely we will derive a decomposition
of $\omega_\beta$  of the form $$
\omega_\beta(A)=\int\mu(\d\om)E_\omega(A) \eqno(2.2)$$ where
$\mu(\d\om)$ is a probability measure on a space of configurations
$\omega$, and for each $\omega$, $E_\omega$ is a linear functional of
$\A_\Gamma$. Of course such an integral decomposition will only be
useful if the structure of the $E_\omega$ is considerably simpler
than $\omega_\beta$ itself and if we are able to analyse the measure
$\mu(\d\omega)$. Below we show that for a certain class of
Hamiltonians there is a quasi-state decomposition to which one can
give a stochastic  geometric interpretation. The resulting
stochastic geometric model has proved to be rather helpful in
revealing structural properties of the states (see Section 3 and 4).

The linear functionals $E_\omega$ will in general not be states and
for infinite systems they are in general only defined on a dense
subalgebra of the form $$ \A_{\rm
loc}=\bigcup_{\Lambda\subset\Gamma}\A_\Lambda \eqno(2.3)$$ where the
union is over finite subsets $\Lambda$ of $\Gamma$. Implicit in
(2.3) is the natural embedding of $\A_\Lambda$ in
$\A_{\Lambda^\prime}$ for $\Lambda\subset\Lambda^\prime$. Although
the $E_\omega$ are typically not positive and not even bounded, they
will be states of an abelian subalgebra of $\A$. We therefore call
these functionals {\it quasi-states\/}.

\headb{2.1}{Poisson integral representations}

In this subsection we consider a quantum spin model defined on a
finite set of sites $\Gamma$, which is completely arbitrary at this
point. The set of ``bonds'', denoted by $\B$, is a collection of
subsets of $\Gamma$. Soon we will specialize to the case where the
bonds $b\in\B$ are pairs of sites and in the case of
antiferromagnetic models it will be important that $\Gamma$ be given
a bipartite struture: $\Gamma =\Gamma_A\cup\Gamma_B$,
$\Gamma_A\cap\Gamma_B=\emptyset$. $\Gamma_A$ and $\Gamma_B$ are
called the $A$- and the $B$-sublattice respectively. In that
situation a bond $b=\{x,y\}\in\B$ will be called {\it
ferromagnetic\/}  if $x$ and $y$ belong to the same sublattice and
{\it antiferromagnetic\/} if $x$ and $y$ belong to different
sublattices. We denote the set of  ferromagnetic bonds by $\B_F$ and
the set of antiferromagnetic bonds by $\B_{AF}$.

A complex Hilbert space $\H_x$ is associated with each site
$x\in\Gamma$. Although this is not necessary for most of what
follows, we will assume that all $\H_x$ have the same finite
dimension $d$: $\H_x\cong\Cx^d$. A spin $S$ system has $d=2S+1$,
$S=\tover12,1,\tover32,\ldots$. The observables at site $x$ are the
elements of $\B(\H_x)\cong M(d,\Cx)$. The Hilbertspace of the system
is then $\H_\Gamma=\bigotimes_{x\in\Gamma}\H_x$ and the algebra of 
observables is $\A_\Gamma=\bigotimes_{x\in\Gamma}M(d,\Cx)$.
Physically important observables are usually expressed in terms of
the spin matrices $S^1,S^2,S^3$ which are the generators of the
$d$-dimensional irreducible unitary representation of SU(2). They
satisfy the commutation relations $$
[S^\alpha,S^\beta]=i\sum_\gamma\epsilon_{\alpha\beta\gamma}S^\gamma
\eqno(2.4)$$ where $\alpha,\beta,\gamma\in\{1,2,3\}$ and
$\epsilon_{\alpha\beta\gamma}$ is the completely antisymmetric
tensor with $\epsilon_{123}=1$. One often combines the three spin
matrices into a vector $\es= (S^1,S^2,S^3)$. The magnitude of this
``spin-vector'' is related to $S$ by $\es\cdot\es=S(S+1)\idty$.

A quantum spin modeled is defined by its Hamiltonian $H=H^*
\in\A_\Gamma$ which we write in the form  $$
H=-\sum_{b\in\B}J_b(h_b-1) \eqno(2.5)$$ with
$h_b\in\A_b\cong\bigotimes_{x\in b}M(d,\Cx)$, and we assume $J_b >0$
for all $b\in\B$. 

To obtain a quasi-state decomposition of $\omega_\beta$, we start
from the following Poisson integral formula: $$ e^{-\beta
H}=\int\rho_\beta^J(d\omega)K(\omega) \eqno(2.6)$$ where:

$\bullet$ $\rho^J_\beta(d\omega)$ is the probability measure of a
product of independent Poisson processes, one for each bond $b\in\B$,
running over the time  interval $[0,\beta]$, and with rates $J_b$. 
For the time being we draw the configurations $\omega$ for this
process  as in Figure 1 (for evident reasons the figures are drawn
for the simplest case of a one-dimensional system:
$\Gamma=[a,b]\in\Ir$ and $\B =\{\{x,x+1\}\mid a\leq x\leq b-1\}$.

$\bullet$ $K(\omega)$ is a product of operators $h_b$, one for each
bond occurring in $\omega$ and ordered according to the times at
which they occur: $K(\omega)=\tprod h_{b_1}h_{b_2}\cdots h_{b_n}$ if
$\omega$ is the set of time-indexed bonds $\{(b_1,t_1),
(b_2,t_2),\ldots,(b_n,t_n)\}$ with $t_1<t_2<\cdots< t_n$.

An important quantity is the partition function $\Z_\beta=\Tr
e^{-\beta H}$, which by (2.6) and linearity of the trace is given by:
$$ \Z_\beta=\int\rho_\beta^J(d\omega)\, \Tr K(\omega) \eqno(2.7)$$
We have found that for a quite large class of interactions $h$  the
following is true:

$\bullet$ $\Tr K(\omega)>0$ for all $\omega$ and this number can be
computed in terms of relatively simple geometric properties of
$\omega$.

$\bullet$ the diagonal matrix elements of the operators $K(\omega)$
are all non-negative in a certain tensor product basis of the
Hilbert space of the system.

Let us consider some elementary examples of this before we proceed.
For a more detailed discussion of these examples we refer to
(Aizenman and Nachtergaele, 1993a).

\vbox{ \vtop to 9truecm{\vfill} \caption{Figure 1. A typical
configuration of the Poisson process $\rho^J_\beta(d\omega)$.}}

\headc{Example 1}

Let $h$ be the operator which interchanges the states of the two
sites, i.e. $h\phi\otimes\psi=\psi\otimes\phi$ for any two vectors
$\phi,\psi \in \Cx^d$. In any basis of $ \Cx^d$ the matrix elements
are 0 or 1 and a  fortiori non-negative. $K(\omega)$ represents a
permutation $\pi(\omega)$ of the sites in $\Gamma$, and its trace is
easily seen to be $d^{\hbox{\eightrm $\#$ cycles in }\pi(\omega)}$.
The number of cycles in $\omega$ becomes a geometric property of the
configuration if we replace the Poisson-``beeps'' by two horizontal
lines that cross each other as in Figure 2. With the convention of
periodic boundary conditions in the vertical direction the number of
cycles in the  permutation $\pi(\omega)$ is then equal to the number
of loops in $\omega$, which we will frequently denote by $l(\omega)$.

With $d=2$ this interaction is equivalent to the usual spin 1/2
Heisenberg ferromagnet: $h_{\{x,y\}}=\tover12 +\tover12\pauli_x
\cdot\pauli_y$, where $\pauli$ is the vector with as its components
the three canonical Pauli-matrices.

\vbox{ \vtop to 9truecm{\vfill} \caption{Figure 2. A typical
configuration of the Poisson process $\rho^J_\beta(d\omega)$
decorated for the ferromagnetic models of Example 1.}}

\headc{Example 2}

For the second example (which includes the spin 1/2 Heisenberg
antiferromagnet) we require $\Gamma$ to be bipartite and all bonds
are assumed to be  antiferromagnetic in the sense of above (so,
$\B=\B_{AF}$). We consider the interaction $h$ defined by $$
h=\sum_{m,m^\prime} (-1)^{m-m^\prime} \vert m,-m\rangle\langle
m^\prime,-m^\prime\vert \eqno(2.8)$$ where $\{\vert m\rangle\}$ is a
basis of $\Cx^d$ given by the eigenvectors of the third component
$S^3$ of the spin with eigenvalues $m$. In the  case $d=2$ this is
the interaction for the standard Heisenberg  antiferromagnet
($h=\tover12 -\tover12\pauli_x\cdot\pauli_y$)but the  same
expressions define an interesting interaction for any finite
dimension  (any magnitude of the spin) and was first studied by
Affleck (Affleck, 1985) and later in (Affleck, 1990; Kl\" umper,
1990; Batchelor and Barber, 1990). The interaction is proportional
to the projection operator onto the singlet vector for a pair of
spins and hence antiferromagnetic. It can be shown that the diagonal
matrix elements  of the corresponding $K(\omega)$ (but not the
off-diagonal ones) are non-negative and there is again a simple
formula for the trace. Each Poisson-``beep'' is now replaced by two
parallel horizontal lines as shown in Figure 3. Again this turns
$\omega$ into a configuration of loops (assuming periodic boundary
conditions in the vertical direction) and $\Tr
K(\omega)=d^{l(\omega)}$.

\vskip 5pt 

The examples given above are not the most general ones that can be
treated but they are in some sense the two basic ones as will become
clear at the end of this section. Unlike their classical analogues,
quantum ferro- and antiferromagnets behave in a very different way
and it is therefore not a surprise that they lead to two very
different  stochastic geometric models.

We complete the quasi-state decomposition of $\omega_\beta$ by
establishing  the relation between expectation values of observables
for the quantum spin system on the one hand and probabilities of
events (or more generally expectations of random variables) in the
probability measure describing the stochastic geometric model on the
other hand. From (2.6) it follows that for any local observable $A$
for the quantum spin chain $$ \omega_\beta(A) \equiv {\Tr A
e^{-\beta H}\over\Tr e^{-\beta H}} =\int\mu(d\omega)E_\omega(A)
\eqno(2.9)$$ where $$
\mu(d\omega)=(\Z_\beta)^{-1}\rho^J_\beta(d\omega)\Tr K(\omega)
\eqno(2.10)$$ and $$ E_\omega(A)={\Tr A K(\omega)\over\Tr K(\omega)}
\eqno(2.11)$$ $\mu(d\omega)$ is a probability measure on the
configurations $\omega$ and for $A$ fixed $E_\omega(A)$ is a random
variable. We found that for many important observables $A$ this
random variable can in fact be given a simple geometric
interpretation. Take e.g. $A=S^3_x S^3_y$. One then finds: $$
E_\omega(S^3_x S^3_y)= \cases{ C(S)  \Ind{\hbox{$(x,0)$ and $(y,0)$
are on the same loop} } &for Example 1\cr (-1)^{\vert x-y\vert} C(S) 
\Ind{\hbox{$(x,0)$ and $(y,0)$ are on the same loop} } &for Example
2\cr} \eqno(2.12)$$ with $C(S)=(\sum_{m=-S}^S m^2)/(2S+1)=S(S+1)/3$
and where $(x,t)\in \Gamma\times[0,\beta]$ denotes a space-time
point. $\Ind{$\phantom{a}\cdot$\phantom{a}}$ denotes the indicator
function of  the event described between the brackets. Hence, the
spin-spin correlation is proportional to the probability, with
respect to the effective probability measure $\mu(d\omega)$ on the
space of  loop configurations, that two sites are on the same loop
of $\omega$:  $$ \langle S^3_x S^3_y \rangle = \cases{ C(S)  
\Probin{\mu}{$(x,0)$ and $(y,0)$ are on the same loop}  &for Example
1\cr
 (-1)^{\vert x-y\vert} C(S)   \Probin{\mu}{$(x,0)$ and $(y,0)$ are
on the same loop} &for Example 2\cr} \eqno(2.13)$$ For $\omega$
fixed (2.11) defines a functional of the algebra of observables
which is a state on a certain abelian subalgebra (see (2.20)). We
therefore call (2.9) a quasi-state decomposition of the state
$\omega_\beta$.

\vbox{ \vtop to 9truecm {\vfill} \caption{Figure 3. A typical
configuration of the Poisson process $\rho^J_\beta(d\omega)$
decorated for the antiferromagnetic models of Example 2.}}

Examples 1 and 2 can be combined in order to obtain a quasi-state
decomposition for a class of models with mixed ferro- and
antiferromagnetic interactions. 

In the statement of the following theorem some extra notation is
used: for any configuration $\sigma=(\sigma_x)_{x\in\Gamma}$ we
denote by  $\vert\sigma\rangle$ the vector in $\H_\Gamma$ given by $$
\vert\sigma\rangle=\bigotimes_{x\in\Gamma} \vert\sigma_x\rangle
\eqno(2.14)$$ where $\vert\sigma_x\rangle$ is the eigenvector of
$S^3_x$ belonging to the  eigenvalue
$\sigma_x\in\{-S,-S+1,\ldots,S\}$. 
$\sigma(t)=(\sigma_x(t))_{x\in\Gamma}$ is a piecewise constant
function on $\Gamma\times[0,\beta]$ taking values in
$\{-S,-S+1,\ldots,S\}$. A configuration of the Poisson process
becomes a configuration of loops if we draw the ferromagnetic bonds
as in Example 1 (Figure 2) and the  antiferomagnetic ones as in
Example 2 (Figure 3). As before, $l(\omega)$ denotes the number of 
loops in $\omega$. We call $\sigma(t)$ {\it consistent\/} with a 
configuration of loops $\omega$ if the discontinuities only occur at
the time-indexed bonds in $\omega$ (but $\omega$ need not be
discontinuous at a bond) and if in addition for each  time-indexed
bond $b=(\{x,y\},t)\in\B\times [0,\beta]$: $$\eqalignno{
\sigma_x(t^-)&=\cases{\sigma_y(t^+)&if $\{x,y\}$ is a ferromagnetic
bond\cr
                      -\sigma_y(t^-)&if $\{x,y\}$ is an
antiferromagnetic bond\cr} &(2.15)\cr
\sigma_x(t^+)&=\cases{\sigma_y(t^-)&if $\{x,y\}$ is a ferromagnetic
bond\cr
                      -\sigma_y(t^+)&if $\{x,y\}$ is an
antiferromagnetic bond\cr} &(2.16)\cr }$$

\thm{2.1} For a Hamiltonian $H$ of the form (2.5) and such that
$h_b$ is the interaction of Example 1 if $b\in\B_F$ and $h_b$ is the
interaction of Example 2 if $b\in\B_{AF}$ we have the following
relations: \nl  i)  $$   \langle \sigma^\prime \mid e^{-\beta H}
\mid \sigma \rangle = \sign(\sigma^\prime,\sigma)\int
\rho^J_\beta(\d\om)   \sum_{\sigma\hbox{\eightrm \  consistent with
}\omega}  \bI[ \sigma(\beta)=\sigma^\prime,  \sigma(0)=\sigma]
\eqno(2.17)$$ where $\sign(\sigma^\prime,\sigma)=(-1)^{N_{AF}(\pi)}$
with $N_{AF}(\pi)$ denoting the number of transpositions $t_{x,y}$
with $\{x,y\}\in\B_{AF}$ in any permutation $\pi$ of the sites in
$\Gamma$ such that $\sigma^\prime_x=\sigma_{\pi(x)}$ (the sign is
well-defined because of the bipartite structure of the model).\nl 
ii) the partition function is given by: $$ \Z_\beta=\Tr e^{-\beta H}=
\int\rho^J_{[0,\beta]}(\d\om) \, (2S+1)^{l(\om)} \eqno(2.18)$$ iii)
the equilibrium expectation values of observables  which are
functions of  the operators $S^3_x$  can be expressed as $$ {\Tr
f((S_x^3)_{x\in\Gamma})e^{-\beta H}\over \Z_\beta} =\int\mu(\d\om)\,
E_\om(f) \eqno(2.19)$$  where $\mu(\d\om)=\Z_\beta^{-1}
\rho_{[0,\beta]}(\d\om) \,  (2S+1)^{l(\om)}$ and the expectation
functional $E_\om(f)$ is obtained  by averaging, with equal weights,
over all the spin configurations consistent with $\omega$: $$
E_\omega(f)={1\over (2S+1)^{l(\omega)}} \sum_{\sigma\hbox{\eightrm \
consistent with }\omega} f(\sigma(t=0)) \eqno(2.20)$$ \endthm

\headb{2.2}{Quasi-state decompositions for more general models}

We now show that the method yielding the quasi-state decompositions
of Theorem 2.1 can be extended, with little effort, to a much larger
class of  Hamiltonians of the form $$
H=-\sum_{b\in\B}h_b=\sum_{\{x,y\}\in\B}\sum_{k=0}^{2s} \tilde
J_k(x,y) (\es_x\cdot\es_y)^k \eqno(2.21)$$ It is in fact useful to
think of the bonds here as a pair $\{x,y\}$ of sites together with a
label $k\in\{1,2S\}$. So, for a nearest neighbour model, we now have
$2S$ bonds associated with each pair of nearest neighbour sites and
also $2S$ independent Poisson processes with rates $J_k(x,y) >0$.
The $J_k(x,y)$ are not the same as the $\tilde J_k(x,y)$ because it
will be necessary to reorganize the $2S$ interaction terms in (2.21)
as follows: $$ \sum_{k=0}^{2s}\tilde J_k(x,y) (\es_x\cdot\es_y)^k
=-\sum_{k=0}^{2s} J_k(x,y) Q_k(\es_x\cdot\es_y) \eqno(2.22)$$ where
$Q_k$ are particular polynomials of degree $k$ to be specified below.
Hence the $J_k$ will be related to the $\tilde J_k$ by a linear
transformation.

The most straigthforward way to obtain a stochastic geometric
representation for a quite extensive class of spin $S$ Hamiltonians
of the form (2.21) is by  representing them as a spin $1/2$ system. 
To this end we introduce the (up to a phase) unique isometry $V^{(S)}
:\Cx^{2S+1}\to(\Cx^2)^{\otimes 2S}$, which intertwines the spin $S$
representation of SU(2) with $2S$ copies of the spin $1/2$
representation: $$ V^{(S)} D^{(S)}(g)=(D^{(1/2)}(g))^{\otimes 2S}
V^{(S)}\ ,\qquad g\in \hbox{ SU(2)} \eqno(2.23)$$ $V^{(S)}$
satisfies: $V^{(S)*}V^{(S)}=\idty$ and $V^{(S)}V^{(S)*}= P^{(S)}$
where $P^{(S)}$ is the orthogonal projection onto the spin $S$
states in the $2S$-fold tensor product of the spin $1/2$ states. The
spin $S$ states coincide with the states that are symmetric under
permutations of the $2S$ factors in $(\Cx^2)^{\otimes 2S}$, and
hence $P^{(S)}$ can be expressed as $$
P^{(S)}\bigotimes_{i=1}^{2S}\phi_i={1\over (2S)!} \sum_{\pi}
\bigotimes_{i=1}^{2S}\phi_{\pi(i)}\ ,\qquad \phi_i\in\Cx^2
\eqno(2.24)$$ It follows by differentation of the intertwining
property (2.23) that the spin $S$ operators $S^\alpha,\alpha=1,2,3$,
can be expressed in terms of $2S$ copies of the Pauli matrices (for
$S=\tover12$, $S^\alpha=\tover12\sigma^\alpha$): $$
 V^{(S)}S^\alpha=\tover12\sum_i^{2S}\sigma^\alpha V^{(S)}
\eqno(2.25)$$ We can now define a class of Hamiltomians $H$ for a
spin $S$ system on a lattice $\Gamma$ by representing it as a spin
$1/2$ model on the extended lattice
$\tilde\Gamma=\Gamma\times\{1,\ldots,2S\}$. With each $x\in\Gamma$
we associate a set of $2S$ sites of the form $(x,k)$ in
$\tilde\Gamma$. The set of bonds $\tilde\B$ is constructed as
follows. For $k=1,\ldots,2S$ and $b=\{x,y\}\in\B$ we define $b_k=
\{(x,1),(y,1),\ldots,(x,k),(y,k)\}$. Then $\tilde\B=\{b_k\mid
b\in\B,1\leq k\leq 2S\}$. The interactions $\tilde h_{b_k}$ are
defined by $$ \tilde h_{b_k}=\prod_{l=1}^k h_{(x,l),(y,l)}
\eqno(2.26)$$ where $h_{(x,l),(y,l)}$ is one of the following two
operators, depending on  whether $b$ is a ferro- or
antiferromagnetic bond: $$ h_{(x,l),(y,l)}=\cases{ \tover12+\tover12
\pauli_{(x,l)}\cdot\pauli_{(y,l)} &if $b$ is ferromagnetic\cr
\tover12-\tover12 \pauli_{(x,l)}\cdot\pauli_{(y,l)} &if $b$ is
antiferromagnetic\cr} \eqno(2.27)$$ It is then straightforward to
derive (as an extension of Theorem 2.1) a  quasi-state decomposition
and the corresponding stochastic geometric representation for a
Hamiltonian of the form $$ H=-V^*\left(\sum_{b\in
\B}\sum_{k=1}^{2S}J_k\tilde h_{b_k}  \right)V \eqno(2.28)$$ where
$V=\bigotimes_{x\in\Gamma}V_x^{(S)}$.  We will describe to resulting
stochastic geometric representation in detail  below. But first we
address the question of how to express a Hamiltonian of the form
(2.28) in terms of the spin $S$ matrices. In particular what are the
polynomials $Q_k$ appearing in (2.22). That (2.28) is equivalent with
(2.22) for some particualar choice of the polynomials $Q_k$ follows
directly form the SU(2) invariance which one can check using the
intertwining relation (2.23). Explicit knowledge of the $Q_k$ is
important if one wants to check whether some particular Hamiltonian
(e.g. one that is discussed in the physics literature) has a
stochastic geometric representation of the kind developed here or
not.

So, we are looking for two sets of polynomials $\{Q^{F}_k\}$ and 
$\{Q^{AF}_k\}$ of degree $\leq 2S$ such that $$\eqalignno{
Q^{F}_k(\es_x\cdot\es_y)&=V^*\prod_{l=1}^kt_{(x,l),(y,l)}V&(2.29)\cr
Q^{AF}_k(\es_x\cdot\es_y)&=V^*\prod_{l=1}^kp_{(x,l),(y,l)}V&(2.30)\cr
}$$ where $$\eqalignno{ t_{(x,l),(y,l)}&=\tover12+\tover12
\pauli_{(x,l)} \cdot\pauli_{(y,l)}&(2.31)\cr
p_{(x,l),(y,l)}&=\tover12-\tover12 \pauli_{(x,l)}
\cdot\pauli_{(y,l)}&(2.32)\cr }$$ The sites $x$ and $y$ in
(3.29)-(3.30) just indicate any two distinct sites and play further
no role. We therefore omit these indices in the discussion that
follows. We also denote by $V$ and $P$ the tensor products of the
isometry $V^{(S)}$ and the projection $P^{(S)}$ over the relevant
sites.

The following two lemma's will be useful. \lem{2.2} For any
$k,l,k^\prime,l^\prime\in\{1,\ldots 2S\}, k\neq k^\prime,l\neq
l^\prime$ the following relations hold: $$\eqalign{
t_{k,l}^2&=\idty\cr Pt_{k,l}t_{k',l}&=Pt_{k,l}\cr
Pt_{k,l}t_{k',l}t_{k,l'}&=Pt_{k,l}\cr Pt_{k,l}P&={1\over
(2s)^2}\sum_{i,j=1}^{2s} Pt_{i,j}\cr} \qquad\eqalign{
p_{k,l}^2&=2p_{k,l}\cr Pp_{k,l}p_{k',l}&=Pp_{k,l}\cr
Pp_{k,l}p_{k',l}p_{k,l'}&=Pp_{k,l}p_{k',l'}\cr Pp_{k,l}P&={1\over
(2s)^2}\sum_{i,j} Pp_{i,j}\cr} $$ \endlem \proof{:} The set of
relations for the $t$'s follows directly from (2.24) and the fact
that $t_{k,l}$ is the transposition interchanging the $k$ th factor
of the $2S$-fold tensor product associated with the first site and
the $l$ th factor of the $2S$-fold tensor product associated with
the second site. The relations for the $p$'s follow from the
relations for the $t$'s by using $$ p_{k,l}=\idty-t_{k,l}
\eqno(2.33)$$ which in turn follows directly from (2.31-32). \QEDD
\endproof

\lem{2.3} The operators $Q^{F}_k\equiv Q^{F}_k(\es_0\cdot\es_1)$ and 
$Q^{AF}_k\equiv Q^{AF}_k(\es_0\cdot\es_1)$ defined in (2.29-30)
satisfy the recursion relations $$\eqalignno{ (2S-k)^2 Q^{F}_{k+1}&=
\{(2S)^2 Q^{F}_1 - 2k(2S-k)\}Q^{F}_k - k^2Q^{F}_{k-1}&(2.34)\cr
(2S-k)^2Q^{AF}_{k+1}&= \{(2S)^2 Q^{AF}_1 -
2k(4S-k+1)\}Q^{AF}_k&(2.35)\cr }$$ for $1\leq k\leq 2S-1$ and  $$
\eqalign{ Q^{F}_0&=\idty\cr Q^{F}_1&={1\over 2}\idty+{1\over
2S^2}\es_0\cdot\es_1\cr} \qquad\qquad\eqalign{ Q^{AF}_0&=\idty\cr
Q^{AF}_1&={1\over 2}\idty-{1\over 2S^2}\es_0\cdot\es_1\cr}
\eqno(2.36)$$ \endlem \proof{:} We derive the recursion relation for
the ferromagnetic case. The relation for the antiferromagnetic
operators can be derived in a similar way. We use the formulas of
Lemma 2.2 and the properties of $V$ and $P$. $$\eqalignno{
Q^{F}_1Q^{F}_k&= V^*t_{1,1}V^*Vt_{1,1}\cdots t_{k,k}V&\cr &={1\over
(2S)^2}\sum_{i,j=1}^{2S}V^*Pt_{i,j}t_{1,1}\cdots t_{k,k}V&(2.37)\cr
}$$ The $(2S)^2$ terms in the RHS of the last equality can be taken
together into three groups of terms which are equal: $(2S-k)^2$
terms where $i>k$ and $j>k$, $k^2$ terms where $i\leq k$ and $j\leq
k$, and the rest which are  $2(k-1)(2S-k+1)$ terms where either
$i\leq k<j$ or $j\leq k<i$.  The result is: $$\eqalign{
Q^{F}_1Q^{F}_k&= {(2S-k)^2\over
(2S)^2}V^*t_{i_1,j_1}t_{i_2,j_2}\cdots t_{i_{k+1},j_{k+1}}V\cr
&\quad +{2k(2S-k+1)\over (2S)^2}V^*t_{i_1,j_1}\cdots t_{i_k,j_k}V\cr
&\quad +{k^2\over (2S)^2}V^*t_{i_1,j_1}\cdots t_{i_{k-1},j_{k-1}}V\cr
}\eqno(2.38)$$ where $\{i_1,i_2,\ldots,i_{k+1}\}$ and
$\{j_1,j_2,\ldots,j_{k+1}\}$ are two sets of distinct indices taken
from $\{1,\ldots, 2S\}$.  This proves the recursion relation for
the  $Q^{F}_k$. The recursion realtion for the $Q^{AF}_k$ is derived
in the same way with one small difference: when $i,j\leq k$ the
cases $i=j$ and $i\neq j$ lead to different terms due to the
relations of Lemma 2.2.

For $k=0$ one has $Q^{F}_0=Q^{AF}_0=V^*V=\idty$. For $k=1$ we use 
the relations (2.25) and (3.31-32): $$\eqalign{
Q^{F}_1&=V^*t_{1,1}V=V^*P{1\over (2S)^2}\sum_{i,j}t_{i,j}V\cr
&=\tover12V^*V+{1\over 2S^2}V^*(\tover12\sum_i\pauli_i)\cdot
(\tover12\sum_j\pauli_j)V\cr &={1\over 2}\idty+{1\over
2S^2}\es_0\cdot\es_1\cr\cr }\eqno(2.39)$$ and
$Q^{AF}_1=\idty-Q^{F}_1$ by (2.33). \QEDD \endproof

In particular this lemma shows that $Q^{F}_k$ and $Q^{AF}_k$ are two
complete independent families of rotation invariant operators on
$\Cx^{2S+1}\otimes\Cx^{2S+1}$ because the recursion relations imply
that  the they are polynomials of degree $k$ in the Heisenberg
interaction $\es_0\cdot\es_1$. The recursion relation can also be
used to obtain  explicit expressions for the polynomials $Q^{F}_k$
and $Q^{AF}_k$.  For $S=1$ one finds: $$\eqalign{ Q_0^{F}&= \idty\cr
Q_1^{F}&= \tover12+\tover12 \es_0\cdot\es_1\cr Q_2^{F}&= -\idty
+\es_0\cdot\es_1+(\es_0\cdot\es_1)^2\cr} \quad\eqalign{
Q_0^{AF}&=\idty\cr Q_1^{AF}&=\tover12-\tover12 \es_0\cdot\es_1\cr
Q_2^{AF}&=-\idty+ (\es_0\cdot\es_1)^2\cr }\eqno(2.40)$$ For later
reference we also give here a general formula for the $Q^{AF}_k$: $$
Q^{AF}_k(z) =2^k\left[{(2S-k)!\over
(2S)!}\right]^2\prod_{l=2S-k+1}^{2S}(z_l-z) \eqno(2.41)$$ where
$z_l=\tover12 l(l+1)- S(S+1)$ are the eigenvalues of
$\es_0\cdot\es_1$.  From (2.41) it immediately follows that
$Q^{AF}_k(\es_0\cdot\es_1)$ is of the form $$
Q^{AF}_k(\es_0\cdot\es_1)=\sum_{l=0}^{2S-k} a_{kl} P^{(l)}
\eqno(2.42)$$ where $P^{(l)}$ is the orthogonal projection onto the
states of total spin $l$ in the tensor product of two spin $S$'s.
One can in fact easily determine the  $a_{kl}$ by computing the
eigenvalue $Q^{AF}_k(\es_0\cdot\es_1)$ takes on a state of total
spin $l$: $$ a_{kl}=2^k\left[{(2S-k)!\over
(2S)!}\right]^2\prod_{m=2S-k+1}^{2S}(z_m-z_l) \eqno(2.43)$$ It
follows immediately that $a_{kl}=0$ if $l>2S-k$ and  that $a_{kl}>0$
if $l\leq 2S-k$. From this observation it also follows that the
class of integer spin antiferromagnetic  Valence Bond Solid (VBS)
chains studied in (Arovas et al., 1988) and (Fannes et al., 1989,
1992) do not alow for a stochastic geometric representation of the
kind discussed here; the measure $\mu$ would not be non-negative. On
the other hand the interactions of VBS models are tuned such that
exact cancelations in the measures occur, ie., positive and negative
constributions in the measure cancel each other in such a way that
some types of fluctuations are greatly reduced. The  simplest
non-trivial example of a VBS model is the AKLT chain,  introduced in
(Affleck et al., 1988). For this model one can derive an extremely
simple quasi-state decomposition for its unique ground state which
reveals beautifully  the structure of the state (Nachtergaele,
1993b). 

We now proceed to describe the stochastic geometric representation
which is the result of the discussion above. We consider a spin $S$
model on a lattice $\Gamma$ with Hamiltonian $$
H=-\sum_{\{x,y\}\in\B_F}\sum_{k=0}^{2S}
J_{b,k}(x,y)Q^{F}_k(\es_x\cdot\es_y)
-\sum_{\{x,y\}\in\B_{AF}}\sum_{k=0}^{2S}
J_{b,k}(x,y)Q^{AF}_k(\es_x\cdot\es_y) \eqno(2.44)$$ It is assumed
that the lattice has a bipartite structure in the sense discussed in
Section 2.1. Let $\rho(\d\omega)$ denote the Poisson process on
$\tilde\B$ as defined in Section 2. As in Examples 1 and 2 of
Section 2 $\Tr K(\omega)$ can be  computed in a geometric way. One
proceeds as follows. Because $V$ is an isometry and the cyclicity of
the trace it is obvious that $$\eqalignno{ \Tr
K(\om)&=\Tr_{\H_\Gamma}\tprod_{b_k\in\omega}V^*\tilde h_{b_k}V&\cr
&=\Tr_{\H_{\tilde\Gamma}}P \tilde K(\omega)&(2.45)\cr }$$ where the
$\tilde h_{b_k}$ are defined in (2.26), $\tprod$ denotes a
time-ordered product, and $$ \tilde
K(\omega)=\tprod_{b_k\in\omega}(\tilde h_{b_k}P) \eqno(2.46)$$ If
there would be no projection operators $P$ in (2.46), it would
immedately follow from Theorem 2.1 that $\Tr
K(\omega)=2^{l(\omega)}$. At each site of the original lattice
$\Gamma$ there would then be $2S$ vertical lines, one for each site
in $\tilde\Gamma$. Each lines carries a spin-$\tover12$ degree of
freedom. The time indexed bonds $b_k$ are graphically represented by
the diagrams shown in Figure 4.

\vbox{ \vtop to 9truecm {\vfill} \caption{Figure 4. The diagrams
representing the interactions $Q_k^F$ and $Q_k^{AF}$. The shaded
boxes represent the permutations $\pi_v$ of the lines in each
vertical segment: the index $i_1$ at the top of the leftmost line 
indicates that this line is identified with the line number $i_1$
under the  shaded box etc.}}

The projection operators $P$ reflect the fact that only states which
are symmetric under permutations of the $2S$ factors are relevant.
From (2.24) it is quite obvious how the inserted projection
operators can be incorporated into the stochastic geometric
representation. It is sufficient to average over all permutations of
the lines between any two time-indexed bonds, i.e. we introduce a
random permutation at each vertical segment in $\omega$. By a {\it
vertical segment\/} $v$ in $\omega$ we mean a maximal time interval
at a site $x\in \Gamma$ such that no bond $b_k=(\{x,y\},k)$ occurs
in $\omega$ during that time interval. For each $\omega$ we define
$\nu_\omega(\d\pi)$ to be the uniform probability measure on
configurations $(\pi_v)_{v\in\om}$ of permutations $\pi_v$ of the
set of $2S$ lines running through the segment $v$. It is then
obvious that $$ \Tr\tilde K(\om)=\int\nu_\omega(\d\pi)
2^{l((\omega,\pi))} \eqno(2.47)$$ It is convenient to define
$\tilde\rho^J_\beta(\d\tilde\omega)
=\rho^J_\beta(\d\omega)\nu_\omega(\d\pi)$ with
$\tilde\omega=(\omega,\pi)$.

\thm{2.4} For a Hamiltonian $H$ of the form (2.44) with all
$J_k(x,y)\geq 0$ for $1\leq k\leq 2S$. Without loss of generality we
may assume that $J_0(x,y)=0$. Then: \nl  i)  $$  \langle \sigma'
\mid e^{-\beta H} \mid \sigma \rangle = \int
\rho^J_\beta(\d\tilde\om)   \sum_{\tilde\sigma\hbox{\eightrm \ 
consistent with }\tilde\omega} 
\sign({\tilde\sigma}^\prime,\tilde\sigma)\bI[
\sum_{k=1}^{2S}\tilde\sigma(\beta)=\sigma^\prime, 
\sum_{k=1}^{2S}\tilde\sigma(0)=\sigma] \eqno(2.48)$$ where
$\tilde\sigma=(\tilde\sigma_{(x,k)})$,  $\tilde\sigma_{(x,k)}\in
\{-\tover12,\tover12\}$ and $\sigma=(\sigma_x)$,  $\sigma_x\in
\{-S,S+1,\ldots, S\}$, and
$\sign({\tilde\sigma}^\prime,\tilde\sigma)$ is the sign defined in
Theorem 2.1 (note that this sign depends only on $\sigma$ and
$\sigma^\prime$.)\nl ii) the partition function is given by: $$
\Z_\beta=\Tr e^{-\beta H}= \int\tilde\rho^J_\beta(\d\tilde\om) \,
2^{l(\tilde\om)} \eqno(2.49)$$ iii) the equilibrium expectation
values of observables  which are functions of  the operators $S^3_x$ 
can be expressed as $$ {\Tr f((S_x^3)_{x\in\Gamma})e^{-\beta H}\over
\Z_\beta} =\int\mu(\d\tilde\om)\, E_{\tilde\om}(f) \eqno(2.50)$$ 
where $\mu(\d\tilde\om)=\Z_\beta^{-1}
\tilde\rho^J_\beta(\d\tilde\om) \,  2^{l(\tilde\om)}$ and the
expectation functional $E_{\tilde\om}(f)$ is obtained by averaging,
with equal weights, over all the spin configurations consistent with
$\omega$: $$ E_{\tilde\omega}(f)={1\over 2^{l(\tilde\omega)}}
\sum_{\sigma\hbox{\eightrm \ consistent with }\tilde\omega}
f(\sum_{k=1}^{2S} \tilde\sigma(t=0)) \eqno(2.51)$$ iv) For a general
local observable $A$ one has $$
\omega_\beta(A)=\int\tilde\rho(\d\tilde\omega)E_{\tilde\omega}(A)
\eqno(2.52)$$ with $$
E_{(\omega,\pi)}(A)=2^{-l((\omega,\pi))}\Tr_{\tilde\Gamma}VAV^*
\tprod_{b_k,v\in\omega} \tilde h_{b_k}\pi_v \eqno(2.53)$$ \endthm

There is a special case of the models (2.44) for which the
stochastic geometric  picture simplifies a lot. They are models on a
bipartite lattice with only antiferromagnetic bonds
($\B_F=\emptyset$) and for which only $J_{2S}>0$ and all other
coupling constants vanish. Then only the diagram $D^{AF}_{2S}$
appears and the $2S$ lines belonging to one site always act in
unison, and therefore can equally well be drawn as one line.
Therefore we will refer to this special cases as the {\it
single-line\/} models as oppposed to the general  {\it multi-line\/}
models. The integral over the permutation variables $\pi_\omega$ 
can be carried out explicitely and one finds $$
\int\nu_\omega(\d\pi) 2^{l((\omega,\pi))}=(2S+1)^{l(\omega)}
\eqno(2.54)$$ where $l(\omega)$ is now the number of loops in the
configuration $\omega$ drawn with a single line at each site and
with the diagram $D_1^{AF}$ at each of the time-indexed bonds as if
it were a spin-$\tover12$ antiferromagnet. Up to trivial constants
the Hamiltonians of this model is given by $$
H=-\sum_{b\in\B_{AF}}P^{(0)}_b \eqno(2.55)$$ where $P^{(0)}_b$
denotes the orthogonal projection onto the singlet state at the bond
$b$. In fact $(2S+1)P^{(0)}_b$ is identical to the interaction $h$
given in Example 2 (2.8), and the quasi-state decomposition for
these models can be derived directly---without reference to the
general multi-line models---as was done in (Aizenman and
Nachtergaele, 1993a).

\heada{3}{Main results for the one-dimensional single-line models}

In this section we only consider one-dimensional single-line models,
i.e.  $\Gamma=[-L,L]\subset\Ir$ and the Hamiltonian is  $$
H=-\sum_{x=-L}^L J_x P^{(0)}_{x,x+1} \eqno(3.1)$$ where all $J_x$
are strictly positive and we will mostly assume that $J_x$ is
independent of $x$ or alternatingly takes on two different values.
For the proofs of the theorems in this section and further details
see (Aizenman and Nachtergaele, 1993a).

One is mainly interested in properties of the  ground states of (3.1)
in the thermodynamic limit ($L\to \infty$). The existence of some
specific thermodynamic limits is guaranteed by Theorem 0 of our
study.

\thm{3.0} For each $L\geq 0$ the Hamiltonian $H_L$ has a unique
ground state $\psi_L$ and if $J_x=J_{x+2}$ for all $x$, the
following two limits exist for  all local observables $A$ (finite
algebraic combinations  of the spin matrices $S_x^i, i=1,2,3$ and
$x\in\Ir$): $$ \langle A\rangle_{\rm even} =\lim_{L\to\infty \atop
L\ \rm even}{\langle\psi_L\mid A\psi_L\rangle\over \langle\psi_L\mid
\psi_L\rangle}\quad,\qquad \langle A\rangle_{\rm odd}
=\lim_{L\to\infty \atop L\ \rm odd}{\langle\psi_L\mid
A\psi_L\rangle\over \langle\psi_L\mid \psi_L\rangle} \eqno(3.2)$$
\endthm

A priori it is not obvious whether $\langle\cdot\rangle_{\rm even}$
and  $\langle\cdot\rangle_{\rm odd}$ represent pure phases or not,
and also these two limits are not necessarily distinct. 

Our main results are the following:

\headc{a) Long range order}

The interaction $-P^{(0)}$ is obviously antiferromagnetic in nature: 
it favours states where nearest neighbour spins are antiparallel. It
is therefore natural to ask whether the ground states posses
long-range antiferromagnetic order in the sense that  $$ \langle
S_0^3S_x^3\rangle\sim (-1)^x m^2\qquad \hbox{for } x \hbox{ large}
\eqno(3.3)$$ with $m\neq 0$. This kind of behaviour is called N\'eel
order. Ignoring the quantum fluctuations such a state could be
pictorially represented by $$
\cdots\up\down\up\down\up\down\up\down\up\down\up\down
\up\down\up\down\up\down\up\down\up\down\up\down\cdots $$ It turns
out that N\'eel order does not occur in the models  with Hamiltonian
(3.1). Absence of N\'eel order is in fact expected to hold for any
one-dimensional quantum antiferromagnet based on the very general
argument of (Mermin and Wagner, 1966). So far a rigorous proof of
the general statement has not been found but for  the family of
models under consideration here we have the following theorem.

\thm{3.1} The infinite volume ground states described in (3.2) do
not have N\'eel order: $$ \lim_{x\to\infty}\langle
S_0^3S_x^3\rangle_{\rm even (odd)}=0 $$ \endthm

The interaction $-P^{(0)}$ could however give rise to another type
of long-range order. Although for a chain of length $\geq3$ there is
no state in which all nearest neighbour pairs of spins are exactly
in the singlet state, there are two states in which  half of the
nearest neighbour pairs do form a singlet: all the spins at an even
site could form a singlet with their nearest neighbour on the right
or, alternatively, with their nearest neighbour on the left. This
leads to two  states of periodicity 2 with respect to lattice
translations. Pictorially these two states would look as follows:

\vskip .5 cm

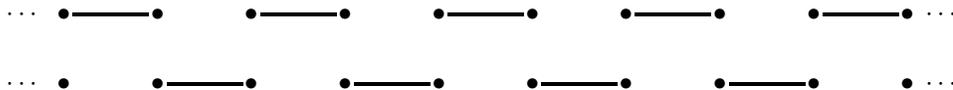
\centerline{$\cdots$ \
\site\bond\site\bondspace\site\bond\site\bondspace
\site\bond\site\bondspace\site\bond\site\bondspace\site\bond\site \
$\cdots$}

\vskip .5 cm

\centerline{$\cdots$ \
\site\bondspace\site\bond\site\bondspace\site\bond\site
\bondspace\site\bond\site\bondspace\site\bond\site\bondspace\site \
$\cdots$}

\vskip .5 cm

This kind of long-range order is called dimerization for obvious
reasons. Again we are ignoring the quantum fluctuations and one can
actually not expect that these are the true ground states. Also in
cases where there  is only a preference of the spins at even sites
to form singlet states with their left or right neighbours, we call
the state (partially) dimerized.

The stochastic geometric representation permits us to prove the
following result related to the Affleck-Lieb dichotomy (Affleck and
Lieb, 1986).

\thm{3.2} For the ground states of the translation invariant  model
(3.1) ($J_x\equiv J$), one of the following holds:

i) either the translation symmetry is spontaneously broken in  the
infinite volume ground states

ii) or the spin-spin correlation function decays slowly
(non-exponential)  with  $$ \sum_x \vert x\langle S^3_0
S^3_x\rangle\vert=+\infty $$

In the first case, the symmetry breaking is manifested in the
non-invariance of the pair correlation: $$ \langle
\es_0\cdot\es_{1}\rangle_{\rm even}\neq \langle
\es_{1}\cdot\es_{2}\rangle_{\rm even} =\langle
\es_{0}\cdot\es_{1}\rangle_{\rm odd} $$ \endthm

The stochastic geometric representation in fact gives us a  detailed
picture of the correlations in the dimerized phase. We find the
following behaviour: in the state where the spins on the even sites
are more correlated with their neighbours to the right one finds
that with probability 1 some spins on the left  half-infinite chain
$(-\infty,x]$ form a singlet with some spins on the right
half-infinite chain $[x+1,+\infty)$, for each even $x$. In the same 
state this probability is $<1$ for odd $x$.

\headc{b) Decay of correlations}

Theorems 3.1 and 3.2 do not specify how fast  $\langle
S_0^3S_x^3\rangle$ converges to zero as $x\to\infty$. The decay rate
by itself is interesting information and in particular one  would
like to show that the decay is exponentially fast (existence of a 
finite correlation length) in the cases where the translation
symmetry is broken. More generally one can consider the truncated
correlation function of any two local observables $A$ and $B$. For
any local observable $C$ we denote by $C_x$ the observable obtained
by translating $C$ over $x$ and let $\supp C$  be the smallest 
interval $[a,b]$ in the chain such that $C$ is localized in 
$[a+1,b-1]$.  We have an estimate for the truncated correlation
function $$ \vert\langle A\, ; B_x\rangle\vert =\langle
AB_x\rangle-\langle A\rangle\langle B_x\rangle \eqno(3.4)$$ in terms
of the truncated two-point function $\tau(x;y)$ of an associated
two-dimensional Potts model. For the translation invariant case the
Potts model is at its self-dual point and for the staggered model it
can be taken to be in the high temperature phase. The precise
definition of this Potts model is given in (Aizenman and
Nachtergaele, 1993a).  The comparison theorem reads as follows.

\thm{3.3} Let $\langle\cdot\rangle$ denote the expectation in the
ground state of a finite chain containing an even number of sites,
or in one of the limiting states $\langle\cdot\rangle_{\rm even}$ or
$\langle\cdot\rangle_{\rm odd}$. Then, for any pair of local
observables $A$ and $B$ of the quantum spin chain there are
constants $C_A$ and $C_B$ such that $$ \vert\langle
A\,;B_x\rangle\vert\leq C_A C_B\sum_{y\in\supp A\atop z\in\supp
B_x}\tau(y\,;z) $$ \endthm     In this theorem $\supp A$ is roughly
equal to the set of sites in the lattice on which the observable $A$
acts non-trivially.

\headc{c) The spectral gap}

One says that the system has a spectral gap of magnitude at least
$\gamma>0$ in an infinite volume ground state $\langle\cdot\rangle$
if for any local observable $A$ the following inequality holds $$
\lim_{L\to\infty}\langle A^*[H_L,A]\rangle\geq \gamma(\langle
A^*A\rangle -\vert\langle A\rangle\vert^2) \eqno(3.5)$$ This is
equivalent to the GNS Hamiltonian having a spectral gap separating
the ground state from the rest of the spectrum. It is therefore
obvious that the existence of a spectral gap would also follow from
an estimate of the type $$ \vert\langle A e^{-tH} B\rangle\vert \leq
C_A C_B e^{-\gamma t} \eqno(3.6)$$ One can actually prove the
analogue of Theorem 3.3 for the observable $B$ being translated in
imaginary time instead of in space and this implies the following
theorem.

\thm{3.4} Whenever the two-dimensional Potts model associated with
the quantum spin chain has an exponentially decaying truncated
two-point function there is a spectral gap in the ground states
$\langle\cdot\rangle_{\rm even}$ and $\langle\cdot\rangle_{\rm odd}$.
\endthm

\heada{4}{Results, open problems, and conjectures for the multi-line
models}

Also here we retrict ouselves to one-dimensional models. Some 
important questions related to higher dimensional models were briefly
discussed in (Nachtergaele, 1993a). We will also only consider
translation invariant models, i.e. the coupling constants $J_k$ do
not depend on the position in the chain. So, the Hamiltonians under
investigation here are $$ H=-\sum_{x=-L}^{L-1}\sum_{k=0}^{2S}J_k
Q^{AF}_k(\es_x\cdot\es_{x+1}) \eqno(4.1)$$ where $Q^{AF}_k$ is the
polynomial of degree $k$ defined in (2.41) and we assume that
$J_k\geq 0$ for $1\leq k\leq 2S$.

\headb{4.1}{The excess spin operator}

In the stochastic geometric picture of the ground state provided by
the Poisson integral representation of Section 2, it is rather clear
intuitively that the quantity $$ \Sh_x^\alpha=\sum_{y>x}S^\alpha_y $$
might indeed make sense when the spin-spin correlations have
sufficient decay. We call these quantities the excess spin
operators. We would like to give meaning to this infinite sum in
such a way that the associated random variable satisfies $$
E_{\tilde\omega}(\Sh^\alpha_x) =\sum_{\hbox{\eightrm loops\ } \gamma
\hbox{\eightrm\ in\ } \omega \atop \hbox{\eightrm intersecting both\
} (-\infty,x] \hbox{\eightrm and\ } (x,+\infty)} \sum_{(y,k)
\hbox{\eightrm on\ } \gamma} \tover12 \sigma^\alpha_{(y,k)}
\eqno(4.2)$$ A crucial property then is that the total third
component of the spin (or any component for that matter) on the
collection of sites where a given loop intersects the $t=0$ axis, is
always identically zero. So, therefore  only these loops that
intersect both $(-\infty,x]$ and $[x+1,+\infty)$, appear in the sum
(4.2). Because of (2.51) decay of the spin-spin correlation function
translates directly into a restriction on the typical size of the
loops in a configuration $\omega$.

We define local approximants $\Sh_x^\alpha(\epsilon)$ for the excess
spin as follows: $$
\Sh_x^\alpha(\epsilon)=\sum_{y>x}e^{-\epsilon\vert
y-x\vert}S^\alpha_y \eqno(4.3)$$ for $\epsilon >0$, $x\in \Ir$, and
$\alpha=1,2,3$. The  $\Sh_x^\alpha(\epsilon)$ are bounded
self-adjoint operators and generate one-parameter unitary groups,
although they are not representations of SU(2). For all $g\in$ SU(2)
we will denote by $\bg$ the vector in  $\Rl^3$ such that in any
strongly continuous unitary representation $U$,
$U(g)=\exp{i\bg\cdot\hat{\bf S}}$, where $\hat{\bf S}$ denotes the
vector of generators of the representation.

For any state $\langle\cdot\rangle$ of the infinite spin chain there
is a  (up to unitary equivalence) unique Hilbert space $\H_{\rm
GNS}$, a vector $\Omega\in \H_{\rm GNS}$ and a representation $\pi$
of the algebra of observables $\A_\Ir$ on  $\H_{\rm GNS}$ such that
$$ \langle A\rangle=\langle\Omega\mid\pi(A)\Omega\rangle \eqno(4.4)$$
for all $A\in\A_\Ir$. This is called the GNS representation of the
state (Bratteli and Robinson, 1981). A natural dense subspace  of
$\H_{\rm GNS}$ is defined by $$ \H_{\rm loc}=\{\pi(A)\Omega\mid A\in
\A_{\rm loc}\} \eqno(4.5)$$ where $\A_{\rm loc}$ is the subalgebra
of $\A$ formed by all finite combinations of the operators
$S_x^\alpha$.

\thm{4.1} Let $\langle\cdot\rangle$ be a ground state  of a 
Hamiltonian of the form (4.1) and denote by $\H_{\rm GNS}$ its GNS
Hilbert space. If  $$ \sum_x\vert x^3\langle S_0^\alpha
S_x^\alpha\rangle\vert<+\infty, \quad\hbox{or for the single-line
models:} \quad\sum_x\vert x\langle S_0^\alpha
S_x^\alpha\rangle\vert<+\infty \eqno(4.6)$$ then, for every $x\in
\Ir$ there is a strongly continous unitary  representation $U_x$ of
SU(2) defined by the limit $$ U_x(g)\psi=\lim_{\epsilon\downarrow
0}e^{i\bg\cdot\pi(\hat{\bf S}_x(\epsilon))}\psi \eqno(4.7)$$ for all
$\psi\in\H_{GNS}$ with the following properties:\nl i) $U_x(g)$
commutes with $\A_{(-\infty,x]}$\nl ii) $U_{x-1}(g)=e^{i\bg\cdot
\pi(\es_x)}U_x(g)$\nl iii) The domain of the generators of $U_x$
contains $\H_{\rm loc}$.\nl \endthm

The proof of this theorem will be given in (Aizenman and
Nachtergaele, 1993b).

\cor{4.2} There are well-defined self-adjoint generators
$\Sh_x^\alpha$ of the representation $U_x$ which are strong limits
of the local approximants (4.3) and they have a common domain for
all $\alpha=1,2,3$ and $x\in \Ir$. On this domain, which contains
$\H_{\rm loc}$, one has the relation $$
\Sh^\alpha_{x-1}=\pi(S^\alpha_x)+\Sh^\alpha_x \eqno(4.8)$$ \endcor

For situations where the excess spin operators exists one can then
apply the following theorem to prove that sufficient decay of
correlations in a ground state of a quantum spin chain with a
Hamiltonian of the form (4.1) and
$S=\tover12,\tover32,\tover52,\ldots$, necessarily exhibits breaking
of the translation invariance.

\thm{4.3} Let $\omega$ be a pure state of the spin $S$ chain with
halfintegral $S$, which is invariant under even translations and
such that the operators $\Sh_x^\alpha$ exist and have the properties
stated in Theorem 4.1 and Corollary 4.2,  then $\omega$ is not
invariant under odd translations. \endthm \proof{:} As
$\Sh^\alpha_x$ is self-adjoint the following spectral projection is
well-defined: $$ P_x\equiv \hbox{ projection onto subspace where }
\Sh^\alpha_x\in \Ir+\tover12 \eqno(4.9)$$ (4.8) and the
$SU(2)$-commutation relations of the excess spin  operators imply
that the $\Sh^\alpha_x$ have spectrum $\subset \tover12 \Ir$ and that
$$ P_{x-1}=\idty-P_x=P_{x+1} \eqno(4.10)$$ and hence $P_x$ commutes
with any local observable: $P_x\in \A_{(-\infty ,x+2k]}^\prime$ for
all $k\in\Ir$. In an irreducible  representation of the observable
algebra this means that $P_x$ is a multiple of the identity and
because it is a projection we conclude that $P_x=\lambda_x\idty$
with $\lambda_x\in\{0,1\}$. By (4.10) this implies $$
\omega(P_x)=\cases{1&for $x$ even (odd)\cr
                   0&for $x$ odd (even)\cr} \eqno(4.11)$$ and
therefore $\omega$ is not invariant under translations over an odd
distance. \QEDD \endproof

\headb{4.2}{A topological index as order parameter}

The configurations of loops in the plane that occur in the Poisson
integral representation of the quantum spin chains (4.1)  have
interesting topological properties which we expect to be of
determining importance for the behaviour of their ground states. We
will be considering configurations of loops in the infinite plane,
which corresponds to studying the bulk of the quantum spin chain in
the  ground state. Afterwards we also make a remark about
topological effects in finite and half-infinite chains.

A possible configuration of loops $\tom$ for a spin-1 chain is shown
in Figure 5, and in general we consider configuartions which are
constructed using the diagrams $D^{AF}_k$, $k=0,\ldots, 2S$.
Obviously one can construct configurations out of these diagrams
that contain infinite lines.  Such configurations are excluded from
the considerations in this section.  More precisely we impose the
following condition.

\thmm{Condition}{4.4} Let $\tom$ be a loop configuration for a
spin-$S$ chain. $\tom$ satisfies the Condition if all loops are
finite and any point $(x,t)$ in the plane is surrounded by only a
finite number of loops. The set of all configurations  satisfying
this condition will be denoted by $\Omega_0$. The probability
measure $\mu(\d\tom)$ satisfies the Condition if
$\Probin{\mu}{$\Omega_0$}=1$. \endthmm

Here, by ``surrounded by a loop'', we mean that any continuous path
connecting $(x,t)$ to infinity must necessarily intersect that loop.

In a translation invariant ground state of the quantum spin chain 
Condition 4.4  will be satisfied  if there is suffcient deacy of the
spin-spin correlation function, e.g. Condition 4.4 is satisfied if $$
\sum_x\vert x\,\omega(\es_0\cdot\es_x)\vert<\infty \eqno(4.12)$$

\vbox{ \vtop to 9truecm {\vfill} \caption{Figure 5. A typical
configuration of loops $\tilde\omega$ for a spin 1 antiferromagnetic
chain. The loops are shown with the canonical orientation with
respect to $x$ which is used in the definition of  the canonical
winding number (4.13).}}

In fact (4.12) implies that the average number of loops surrounding
any given point is finite: $$\eqalign{ \sum_{x\geq 0}\vert
x\,\omega(S_0^3S_x^3)\vert &=\sum_{x\leq 0, y>0}\vert \omega(
S_0^3S_x^3)\vert\cr &\geq\tover14\sum_{x\leq 0, y>0}\Prob{$x$ is
connected by a loop to $y$}\cr &\geq\tover14\E(\hbox{ connected
pairs }x\leq 0, y>0)\cr &\geq\tover14\E(\hbox{ loops } \gamma \hbox{
surrounding\ } (\tover12,0))\cr }$$

Let $\tom$ be a configuration for which Condition 4.4 is satisfied.
For any $(x,t)\in\Rl^2$ such that $(x,t)$ does not belong to a loop
in $\tom$, we define the canonical orientation of the loops
$\gamma\in\tom$ as the one for which the vertical lines located at
$z\in\Ir$ are oriented in the positive time direction if
$\sign(x-z)(-1)^{[x-z]}=1$ ($[a]$ denotes the integral part of $a$),
and in the negative time direction if $\sign(x-z)(-1)^{[x-z]}=-1$
(see Figure 5). We then define the {\it canonical winding number\/}
$w_{(x,t)}(\tom)$ of $\tom$ with respect to $(x,t)$ by $$
w_{(x,t)}(\tom)=\sum_{\gamma\in\tom} {1\over 2\pi}\int_\gamma
d\theta_{(x,t)} \eqno(4.13)$$ where $\theta$ is defined in Figure 5.
By Condition 4.4 only a finite number of terms in the sum are
non-vanishing and of course $w_{(x,t)}(\tom)$ is an integer. The
following crucial properties, which are straightforward to derive
from the  definition of the canonical winding number, show that the
value of the winding number  is actually a property of $\tom$, i.e.
it does not depend upon the point $(x,t)$ with respect to which it
is computed, except for a trival dependence on the  parity of $x$: $$
w_{(x,t)}(\tom)=w_{(x,s)}(\tom)\equiv A_x(\tom) \eqno(4.14)$$ and  $$
A_{x+1}(\tom)=2S-A_x(\tom) \eqno(4.15)$$ for all
$x\in\Rl\setminus\Ir$ and $s,t\in\Rl$.

What we have found is an index $A_0(\tom)$ which takes on integer
values and which characterizes a global topological property of
$\tom$. It follows that for any  $\Ir$-ergodic probability measure
$\mu(d\tom)$ on $\Omega_0$, $A_0(\tom)$ will take a fixed value with
probability 1 with respect to $\mu(d\tom)$. We claim that the
dimerization transition observed in certain antiferromagnetic
quantum spin chains can be interpreted as a transition between
phases with a different value for the index $A_0(\tom)$. It can be
shown that for each integer $z$ there are indeed  $\tom\in\Omega_0$
for which $A_0(\tom)=z$, with the exception of the case 
$s=\tover12$, where $A_0(\tom)$ can only be $0$ or $1$. But we
expect that only relatively small values will occur in the ground
states of the Hamiltonians (4.1) although we have not been able to
prove this so far. A second important  open problem is to show that
the winding number, which under certain conditions  on the decay of
spin-spin correlations is certainly measurable with respect to
$\mu(\d\tom)$, is also observable for the quantum spin chain. It is
indeed not trivial that there exists an observable $A\in\A$ of the
quantum  spin chain of which the expectation value $\omega(A)$ would
be equal to (or determine uniquely) the value of the topological
index $A_0$ in the probability measure $\mu(\d\tom)$ related to
$\omega$ by (2.52) (with $\beta=+\infty$).

My main interest in the problems mentioned in the previous paragraph
stems from the implications they have for the general structure of
the ground state phase diagram of antiferromagnetic quantum spin
chains. (Affleck and Haldane, 1987) make an thorough analyis of that
phase diagram on the basis of an identifaction of the low-energy
spectrum of spin chains with certain conformal quantum field
theories. Their (non-rigorous) arguments lead to some very
interesting predictions which we can summarize as follows. For the
parameters in the Hamiltonian taking values in the complement of
some submanifold of less than maximal dimension one expects that: i)
for the half-integer spin antiferromagnetic chains the ground states
either exhibit one of a finite number of possible types of
dimerization, in which case they have exponential decay and a
spectral gap, or the ground state is unique and translation
invariant with slow (non-exponential) decay of correlations and  no
spectral gap; ii) for the integer spin antiferromagnetic chains
there also is a finite number of possible types of dimerization and
there is also a translation invariant phase but one with 
exponential decay and a spectral gap (the so-called  Haldane-phase).
So, apart from special choices of the coupling constants where one
is on the boundary between two or more different phases, one expects
that the integer spin chains  always have expontential decay and a
spectral gap where for the half-integer spin chains a gapless
translation invariant phase exists. Our Theorem 4.3 (combined with
Theorem 4.1 and Corollary 4.3) shows that for half-integer spin
chains with antiferromagnetic interactions of the form  (4.1) indeed
there is no translation invariant phse with rapid decay of
correlations. We believe that the different possible phases can be
characterized by the value of the index $A_0$. Note that (4.15) has
no translation invariant solutions for half integer $S$ but that
there is a translation invariant solution $A_x\equiv S$ for integer
$S$. It follows that the gapless translation invariant phase for
half-integer spin chains must have a slowly decaying spin-spin
correlation  function so as to make (4.6) divergent.

The existence of the excess spin operators also allows for a natural
interpretation of the ``exotic'' edge states in finite and
half-infinite integer spin chains. It was observed in Electron Spin
Resonance experiments on the quasi-one  dimensional compound NENP
(Hagiwara et al., 1990,1992; Hagiwara and  Katsumata, 1992) that the
edges interact with a magnetic moment near to it as if they carry
spin 1/2, although the system in all other  respects seems to be
modelled very well by a spin 1 antiferromagnetic chain. It was noted
in (Hagiwara et al., 1990) that if the ground state of this system
resembles the one of the AKLT model (for which the exact ground
state is given in (Affleck et al., 1987,1988)), the spin 1/2 nature
of the edge states could be easily explained. Our proof of the
existence of the excess spin operator is valid for a quite general
class of Hamiltonians. If one assumes that it is correct that a
unique translation invariant ground state for a spin 1 chain implies
that the index $A_x$ is identically equal to $1$ (or equal to $S$
for higher integer spin), it  follows that the excess spin is
half-integral and that the edge states should have half-integral
spin as well. 

\headc{Acknowledgement.} The research discussed in this paper was
carried out in collaboration with Michael Aizenman. 

% -------  T E X T  E N D S  -----------------------------

\refs%-----------------R E F E R E N C E S----------------

\ref\rauthor{Affleck, I.}
\ryear{1985}
\rtitle{Large-$n$ limit of SU($n$) quantum ``spin'' chains.}
\rjournal{Phys. Rev. Lett.}
\rvolume{54}
\rjpp{966--969}
\endref{\journal}

\ref\rauthor{Affleck, I.}
\ryear{1990}
\rtitle{Exact results on the dimerization transition in SU(n)
antiferromagnetic chains.}
\rjournal{J. Phys. C: Condens. Matter}
\rvolume{2}
\rjpp{405--415}
\endref{\journal}

\ref\rauthor{Affleck, I and Haldane, F.D.M.}
\ryear{1987}
\rtitle{Critical theory of quantum spin chains.}
\rjournal{Phys. Rev.}
\rvolume{B36}
\rjpp{5291--5300}
\endref{\journal}

\ref\rauthor{Affleck, I., Kennedy, T., Lieb, E.H., and Tasaki, H.}
\ryear{1987}
\rtitle{Rigorous results on valence-bond ground states in 
antiferromagnets.}
\rjournal{Phys. Rev. Lett.}
\rvolume{59}
\rjpp{799--802}
\endref{\journal}

\ref\rauthor{Affleck, I., Kennedy, T., Lieb, E.H., and Tasaki, H.}
\ryear{1988}
\rtitle{Valence bond solid states in isotropic quantum 
antiferromagnets.}
\rjournal{Comm. Math. Phys.}
\rvolume{115}
\rjpp{477--528}
\endref{\journal}

\ref\rauthor{Affleck, I. and Lieb, E.H.}
\ryear{1986}
\rtitle{A proof of part of Haldane's conjecture on quantum 
spin chains}
\rjournal{Lett. Math. Phys.}
\rvolume{12}
\rjpp{57--69}
\endref{\journal}

\ref\rauthor{Aizenman, M. and Nachtergaele, B.}
\ryear{1993a}
\rtitle{Geometric aspects of quantum spin states.}
\rpublisher{to appear in Comm. Math. Phys}
\endref{\book}

\ref\rauthor{Aizenman, M. and Nachtergaele, B.}
\ryear{1993b}
\rtitle{Geometric aspects of quantum spin states II.}
\rpublisher{in preparation}
\endref{\book}

\ref\rauthor{Arovas, D.P., Auerbach, A., and Haldane, F,D.M.}
\ryear{1988}
\rtitle{Exenteded Heisenberg models of antiferromagnetism: 
analogies to 
the fractional quantum Hall effect}
\rjournal{Phys. Rev. Lett.}
\rvolume{60}
\rjpp{531--534}
\endref{\journal}

\ref\rauthor{Batchelor, M.T. and Barber, M.}
\ryear{1990}
\rtitle{Spin-s quantum chains and Temperley-Lieb algebras.}
\rjournal{J.~Phys. A: Math. Gen.}
\rvolume{23}
\rjpp{L15--L21}
\endref{\journal}

\ref\rauthor{Bratteli, O. and Robinson, D.W.}
\ryear{1979}
\rbook{Operator algebras and quantum statistical mechanics I}
\rpublisher{Springer, New York-Heidelberg-Berlin}
\endref{\book}

\ref\rauthor{Fannes, M., Nachtergaele, B., and Werner,  R.F.}
\ryear{1989}
\rtitle{Exact Antiferromagnetic Ground States for Quantum Chains.}
\rjournal{Europhys. Lett.}
\rvolume{10}
\rjpp{633-637}
\endref{\journal}

\ref\rauthor{Fannes, M., Nachtergaele, B., and Werner,  R.F.}
\ryear{1992}
\rtitle{Finitely correlated states for quantum spin chains.}
\rjournal{Comm. Math. Phys.}
\rvolume{144}
\rjpp{443--490}
\endref{\journal}

\ref\rauthor{Hagiwara, M., Katsumata, K., Renard, J.P., and 
Affleck, I., Halperin, B.I.}
\ryear{1990}
\rtitle{Observation of $S=\tover12$ degrees of freedom in an 
$S=1$ linear
chain Heisenberg antiferromagnet.}
\rjournal{Phys. Rev. Lett.}
\rvolume{65}
\rjpp{3181--3184}
\endref{\journal}

\ref\rauthor{Hagiwara, M. and Katsumata, K.}
\ryear{1992}
\rtitle{Observation of $S=\tover12$ degrees of freedom in an 
undoped $S=1$ linear
chain Heisenberg antiferromagnet.}
\rjournal{J. Phys. Soc. Japan.}
\rvolume{61}
\rjpp{1481--1484}
\endref{\journal}

\ref\rauthor{Hagiwara, M., Katsumata, K., Affleck, I., Halperin,
 B.I., and 
Renard, J.P.}
\ryear{1992}
\rtitle{Hyperfine structure due to the $S=\tover12$ degrees of 
freedom in 
an $S=1$ linear chain antiferromagnet.}
\rjournal{J. Mag. Mag. Materials}
\rvolume{104-107}
\rjpp{839--840}
\endref{\journal}

\ref\rauthor{Haldane, F.D.M.}
\ryear{1983}
\rtitle{Continuum dynamics of the 1-D Heisenberg antiferromagnet:
identification with the O(3) nonlinear sigma model.}
\rjournal{Phys.Lett.}
\rvolume{93A}
\rjpp{464--468}
\endref{\journal}

\ref\rauthor{Kl\"umper, A.}
\ryear{1990}
\rtitle{The spectra of q-state vertex models and related 
antiferromagnetic quantum spin chains.}
\rjournal{J.Phys. A: Math. Gen.}
\rvolume{23}
\rjpp{809--823}
\endref{\journal}

\ref\rauthor{Mermin, N.D. and Wagner, H.}
\ryear{1966}
\rtitle{Absence of ferromagnetism or antiferromagnetism
in one- or two-dimensional isotropic Heisenberg models.}
\rjournal{Phys. Rev. Lett.}
\rvolume{17}
\rjpp{1133-1136}
\endref{\journal}

\ref\rauthor{Nachtergaele, B.}
\ryear{1993a}
\rtitle{A stochastic geometric approach to quantum spin systems.}
\inbook{Probability theory of spatial disorder and phase transition.}
\rauthor{G.R. Grimmett (Ed).}
\rpublisher{Kluwer (to be published, 1994)}
\endref{\inbook}

\ref\rauthor{Nachtergaele, B.}
\ryear{1993b}
\rtitle{A note on quasi-state decompositions of finitely
correlated states.}
\rpublisher{unpublished}
\endref{\book}

\bye
\end